\documentstyle[multicol,aps,epsf,rotate]{revtex}

\begin{document}
\input psfig.tex

\title{Energy Barriers to Motion of Flux Lines in Random Media}
\author{Lev V. Mikheev}
\address{Nordita, Blegdamsvej 17, DK-2100, Copenhagen \O, Denmark}
\author{Barbara Drossel and Mehran Kardar}
\address{Department of Physics, Massachusetts Institute of
Technology, Cambridge, Massachusetts 02139}
\date{\today}
\maketitle
\begin{abstract}
We propose algorithms for determining both lower and upper bounds
for the energy barriers encountered by a flux line in moving through
a two-dimensional random potential. Analytical arguments, supported by
numerical simulations, suggest that these bounds scale
with the length $t$ of the line as $t^{1/3}$ and $t^{1/3}\sqrt{\ln
t}$, respectively.
This provides the first confirmation of the hypothesis that barriers
have the same
scaling as the fluctuation in the free energy.
\end{abstract}
\pacs{PACS numbers: 74.60.Ge, 05.70.Ln, 05.40.+j}
\begin{multicols}{2}
Glassy states are characterized by a complex energy landscape with
many
(metastable) free energy minima\cite{glass1,glass2}. Some commonly
encountered
examples are spin glasses, pinned flux lines in superconductors,
polymers, domain walls in random field and random bond Ising models,
and many more. Typically, the fluctuations between free energy minima
in these systems (either in different realizations, or in the same
random system) scale with observation size $L$ as $L^\theta$. On the
other hand,
dynamic response of the system is limited by the barriers in
free energy encountered in crossing from one minimum to another.  It
is likely
that the scale of these barriers also grows with size as $L^\psi$.
The simplest
assumption is that the two energy scales are comparable, i.e.
$\psi=\theta$.
However, it is also quite possible that the heights of the ridges in
the random
energy landscape scale differently from those of the valleys that
they separate,
with $\psi>\theta$.  Yet another scenario is that transport occurs
mainly along a percolating channel of exceptionally low energy
valleys with $\psi<\theta$.
In this paper we examine a specific glassy system, a flux line moving
in a two
dimensional random potential, for which we demonstrate $\psi=\theta$.

The system we study is inspired by measurements of nonlinear
current-voltage $(I,V)$
characteristics, $V\propto\exp(-\mbox{const}\ I^{-\mu})$, of
disordered  superconductors
in a magnetic field\cite{expreview}. There is emerging consensus that
such behavior
is best described in terms of low-temperature glass
phases\cite{FisherFisherHuse,BoseGlass}: In the weak current regime,
the
dynamical properties are\cite{FisherFisherHuse,JoffeVinokur}
dominated by
activated processes corresponding to bundles of flux lines overcoming
pinning barriers.
The principal difference from the classical picture of flux line
motion
\cite{KimAnderson} is the power-law growth of  barrier energies
$E_B\propto L^{\psi}$,
with  bundle size $L$. The latter in
turn diverges as the superconducting current driving flux flow
decreases,
leading to nonlinear $(I,V)$ dependence quoted above.
As accurate values of critical exponents characterizing vortex glass
phases
(including $\psi$) are presently not available, direct experimental
confirmation or
refutation of the various theoretical models is difficult.
The point of view taken in many of the pioneering papers
\cite{HuseHenley,JoffeVinokur} on the subject  is that $\psi=\theta$.
Here
we shall attempt to put this assumption on a firmer basis.

We shall examine the configurations of a single flux line (FL) in a
random potential
landscape. Equilibrium properties of this system have been
extensively
studied in the context of {\it directed polymers in random media}
(DPRM)
\cite{KardarRev}.
It is known that the FL is pinned by impurities into a glassy state.
Furthermore, by using a transfer matrix method, properties of this
state can
be probed numerically in polynomial time in the line length $t$.
In two dimensions, e.g. for a FL trapped between two copper oxide
planes of
a high-$T_c$ superconductor, some analytical information is also
available.
For example, the fluctuations in the free energy at finite
temperature scale as $t^{1/3}$. Since the scaling behavior of the
pinned FL is governed by a zero-temperature fixed point
\cite{HuseHenley}, energy fluctuations scale in the same way. The
availability
of such results and techniques make this system ideal for
investigation of
barrier energies.

The precise description of the model is as follows: The FL is
discretized
to lie on the bonds of a square lattice, directed along its
diagonal. Each segment of the line can proceed along
one of two directions, leading to a total of $2^t$ configurations
after $t$
steps. These configurations are labelled by the set of integers $
\left\{ x(\tau)
\right\}$ for $\tau=0,1,\cdots,t$, giving the transverse coordinate
of the line
at each step (clearly constrained such that $x(\tau+1)=x(\tau)\pm
1$). To each
bond on the lattice is assigned a (quenched) random energy equally
distributed
between 0 and 1. The energy of each configuration is the sum of all
random bond energies on the line. Without loss of generality, we set
$x(0)=0$.
For each endpoint $(t,x)$ with $x = -t, -t + 2, \cdots, t$, there is
a configuration of minimal
energy $E_{min}(x|t)$ which can be obtained numerically in a time of
order $t^2$.
It is known that for $|x| < x_c \propto t^{2/3}$ the function
$E_{min}(x|t)$ behaves
as a random walk and is thus asymptotically Gaussian
distributed\cite{KardarRev,HuseHenley}. We next examine the energy
barrier that
has to be overcome when  the line is moved from an initial minimal
energy
configuration between $(0,0)$ and $(t,-x_f)$ to a final one between
$(0,0)$ and
$(t,+x_f)$.

The only elementary move allowed is flipping a kink along the line
from one side to the other (except at the end point). Thus the point
$(\tau, x)$ can
be shifted to $(\tau, x \pm 2)$. Each route from the initial to the
final configuration is
obtained by a sequence of such elementary moves. For each sequence,
there
is an intermediate configuration of maximum energy, and a barrier
which
is the difference between this maximum and the initial energy. In a
system at
temperature  $T$, the probability that the FL  chooses a sequence
which crosses a barrier of height $E_B$ is proportional to
$\exp(-E_B/T)$, multiplied by the number of such sequences. We assume
that, as is the case for the equilibrium DPRM, the ``entropic"
factor of the number of paths does not modify scaling behavior. Thus
at sufficiently low temperatures the FL chooses the optimal sequence
which has to overcome the least energy, and the overall barrier is
the minimum
of barrier energies of all sequences.

Since the number of elementary moves
scales roughly as the area between the initial and final lines, the
number of possible
sequences grows as $t^{xt}$. This exponential growth makes
it practically impossible to find the barrier by examining all
possible sequences,
hampering a systematic examination of barrier energies. Rather than
finding
the true barrier energy, we proceed by placing upper and lower bounds
on it.
The lower bound was given in  ref.\cite{HwaFisher}, and scales as
$t^{1/3}$.
In this paper, we present an algorithm for obtaining an upper bound.
Analytical
arguments, suggest that this upper bound grows as $t^{1/3}\sqrt{\ln
t}$,
thus establishing $\psi=1/3$. Since these arguments do not constitute
a
rigorous proof, we verify their validity by numerical simulations.
Computer time and
memory requirements for the construction of this upper bound are
happily
polynomial in $t$.

\begin{figure}
\narrowtext
\centerline{\rotate[r]{\epsfysize=3in
\epsffile{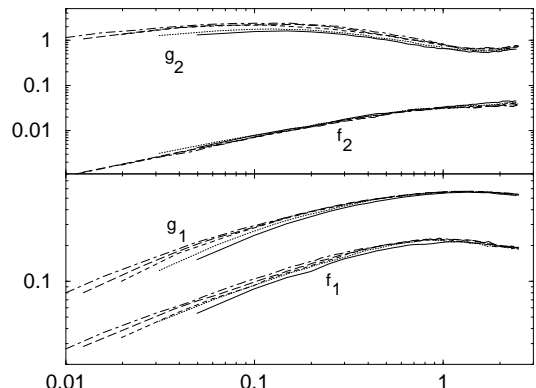}}}
\caption{Scaling functions $f_1(y)$, $f_2(y)$, $g_1(y)$, and $g_2(y)$
(see eqs.(1), (4),
and (5)) for the mean and variance of the
lower and upper bounds; averaged over 2000 realizations of
randomness,
for $t = 256$ (solid), $t = 512$ (dotted), $t = 1024$ (dashed),
$t = 2048$ (long dashed), and $t = 4096$ (dot--dashed).
}
\label{fig1}
\end{figure}

A lower bound for the barrier energy is obtained as
follows\cite{HwaFisher}: Since
the endpoint  of the path has to visit all sites $(t,x)$ with $|x|
\leq x_f$, and since the
energy of  any path ending at $(t,x)$ is at least as large as
$E_{min}(x|t)$, the barrier
energy cannot be smaller than $\max[E_{min}(x|t) - E_{min}(-x_f|t)]$
for ${x \in [-x_f, x_f]}$.
When $x_f$ is sufficiently small, the probability distribution of
this lower bound
is identical to that of the maximal deviation of a random
walk of length $x_f$. The latter is a Gaussian distribution with a
mean value
$\propto \sqrt{x_f}$ and a variance $\propto x_f$. This growth
saturates for
$x_f$ of the order of $t^{2/3}$, leading to the scaling behaviors,
\begin{eqnarray}
  \left\langle {E_-} (t,x) \right\rangle  & = & t^{1/3} f_1(x /
t^{2/3}),\qquad \text{ and}
\nonumber \\
 {\rm var}(E_-)& = & t^{2/3} f_2(x / t^{2/3}) ,
\end{eqnarray}
for the lower bound and its variance.
The functions $f_1(y)$ and $f_2(y)$ are proportional to $\sqrt{y}$
and $y$ for small $y$, respectively,  and go to a
constant for $y = O(1)$. Our simulation results for systems with $t=$
256, 512,
1024, 2048, and 4096 confirm this expectation. Fig.~\ref{fig1} shows
the scaling functions $f_1(y)$ and $f_2(y)$ for different $t$, and
the collapse is quite
satisfactory. However, the initial growth  $\propto \sqrt{x_f}$, is
not clearly seen at these sizes.

To obtain an upper bound for the barrier, we specify an explicit
algorithm for
moving the line from its initial to its final configuration. This is
achieved by
finding a sequence of intermediate steps. It is certainly
advantageous to keep the
intermediate paths as close to minimal configurations as possible. We
first attempt
to move the path in steps from a minimal configuration with endpoint
at $(t, x)$
to  one with endpoint at $(t, x + 2)$, starting at $(t, -x_f)$ and
ending at $(t, x_f)$.
At each step, we obtain a local barrier path which separates two
neighboring
minimal paths. The overall barrier is of course the one with the
highest energy.
While it may occasionally be possible to go from one optimal path to
a
neighboring one in a single elementary move (as defined above), this
is
generally not the case. Minimal paths with neighboring endpoints may
be
quite far apart  at coordinates $\tau < t$. The reason is simple:
suppose the
random potential has a large positive fluctuation, a ``mountain''.
The minimal
energy paths will then circumvent this region by going to its right
or left. The last
path going to the left and the first one going to the right have
almost the same
energy (these energies are strictly equal in the continuum limit).
They form a
{\em loop} which  can be quite large and  is likely to enclose the
barrier when
both paths separate already at small $\tau$. Such loops have been
conjectured \cite{JoffeVinokur,FisherFisherHuse} to play an important
role
 in the low-temperature dynamics of the DPRM. Since the transverse
fluctuations of a minimal path of length $t$ grow as $t^{2/3}$, we
expect the
lateral size of these loops to also be of this order.

\begin{figure}
\narrowtext
\centerline{{\epsfysize=3in
\epsffile{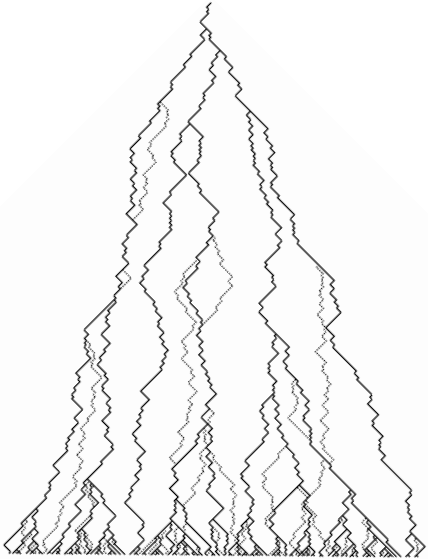}}}
\caption{
Minimal paths of length $t = 256$ to endpoints between
$x = -96$ and  $x = +96$ (solid),
and the barrier paths between them (dotted).
}
\label{fig2}
\end{figure}

The algorithm for moving a line of length $t = 2^n$ from a minimal
configuration $\{x_1(\tau)\}$  to another
one $\{x_2(\tau)\}$, with $x_2(t) = x_1(t) + 2$ is as follows:
If $x_2(\tau) \le x_1(\tau) + 2$ for all $\tau$, we can choose a
sequence of
elementary moves such that at most two bonds of the line are not on
one
or the other minimal path, leading to a barrier of order 1 between
the two.
If $x_2(\tau) > x_1(\tau) + 2$ for some $\tau$, the two paths enclose
a loop. We then
consider the midway points ${(t / 2, x)}$ with $x_1(t / 2) < x <
x_2(t / 2)$.  For each of
these points, we find two minimal segments of length $t / 2$
connecting on one side
to $(0,0)$ and on the other to either $(t, x_1(t))$ or $(t, x_2(t))$
\cite{footnote1}.
The two segments form an almost minimal path of length $t$,
constrained to go through
the point $(t / 2, x)$.  We next move the line  $\{x_1(\tau)\}$
stepwise through
this sequence of almost minimal paths. At each step we first attempt
to move the upper
segment and then the lower one\cite{footnote2}.
The prescription for moving these segments of length $t / 2$  is
exactly the same as for
paths of length $t$: If the distance between two consecutive
configurations is larger than
2 for some  $\tau \in [0, t / 2]$, we consider the points at ${(t /
4, x)}$  in between the
two, and find minimal paths of length $t / 4$ connecting them to the
initial and final
points. Next we attempt to move segments of length $t / 2$ by
repeatedly moving
line portions of length $t / 4$. In some cases, when the energy
barrier is high, it
is necessary to proceed with this construction until the cutoff scale
$t / 2^n = 1$ is
reached. Thus, at each intermediate configuration, the line is
composed of one
minimal segment of length $t/2$, one of length $t/4$, etc; ending
with two smallest pieces
of  length $t / 2^m$ (equal to 1 in the worst case). The barrier
paths resulting from
this construction, and the minimal paths separated by them, are shown
in Fig.~\ref{fig2}.

We now estimate the barrier energy resulting from the above
construction.
Each intermediate path is composed of segments of minimal paths with
constrained
endpoints, and we would like to find the probability distribution for
the highest energy.
Constraining the endpoints of a minimal path of length $\tau$
typically increases
its energy by $E_-(\tau)\propto \tau^{1/3}$. A subset of these
intermediate paths
(those that cross the largest mountains) have constraints imposed on
segments
of length $t$, $t/2$, $t/4$, and all the way down to unity.  The
number of paths in this
subset (henceforth referred to as {\it candidate} barriers) grows as
$N_c(t)\propto t^\alpha$, with $1<\alpha<1+2/3$.
The lower limit comes from noting that at each bisection of a loop,
several new
large loops  are generated, at least one in the upper and one in the
lower half
of the parent loop, thus $N_c\geq t$. The upper limit comes from the
total number of
intermediate steps that grows as $tx_f$. The energy of each candidate
barrier
path is obtained in a manner similar to that of the lower bound:
Instead of finding
the maximum of a random walk of length $x_f\propto t^{2/3}$, we now
have to examine
the sum of the maxima for a sequence of shorter and shorter random
walks added
together. The mean value of this sum is related to the convergent
series,
\begin{eqnarray}\label{meanEc}
\left\langle E_c(t) \right\rangle=\left\langle
E_-(t)+E_-(t/2)+E_-(t/4)+\cdots \right\rangle
&=& \nonumber\\
\left\langle E_-(t) \right\rangle\left(1 + 2^{-1/3} + 2^{-2/3} +
\cdots\right) +A
&\simeq &\nonumber \\
\left\langle E_-(t)\right\rangle\left(1-2^{-{1/3}}\right)^{-1} + A=
4.85... \left\langle E_-(t) \right\rangle&+&A.
\end{eqnarray}
The constant $A$ in eq.(\ref{meanEc}) accounts for the
breakdown of the scaling form of the energy increase  for small
loops.
The mean angle of the smallest loops (of size $t_m = t / 2^m$)
approaches the $45^\circ$
limit; their mean energy growing as $0.5 t_m$. For the larger loops,
the angle
$t_m^{2/3} / t_m$ is small and the energy is $0.23 t_m$. A finite
value of $m$
acts as a cutoff separating the two limits.  The energy difference
per unit length between
small and large paths then leads to the additive constant $A$ (of the
order of unity)
in eq.(\ref{meanEc}).

The barrier energy is the maximum of the $N_c(t)$ energies of all
candidate
barriers. To find its characteristics, we need the whole probability
distribution for
the energy $E_c(t)$. Since $E_c$ is the sum of energies coming from
its minimal
segments, the simplest {\it assumption} is to regard the segment
energies as
{\it independent}, approximately gaussian, random variables. We then
conclude that $E_c(t)$ is also gaussian distributed with a variance,
\begin{eqnarray}\label{varEc}
{\rm var}\left( E_c(t) \right)&=&{\rm var}\left( E_-(t)\right)+{\rm
var}\left( E_-(t/2)\right)+
\cdots\nonumber\\
&\simeq&2.70... {\rm var}\left( E_-(t) \right) \propto t^{2/3}.
\end{eqnarray}
Since the different segments are in fact constructed through a
specific recursive
procedure, their independence cannot be justified. Thus the
statements of the
gaussian nature of $E_c(t)$, and the variance in eq.(\ref{varEc}),
should be
regarded as plausible assumptions that appear to be supported by the
numerical
simulations.

It can be checked easily that (for large $N$), the maximum of $N$
{\it independent}
gaussian variables of mean $a$ and variance $\sigma^2$, is a gaussian
of mean
$a+\sigma\sqrt{2\ln N}$ and variance $\sigma^2/(2\ln N)$.  Since the
$N_c(t)$
candidate barriers have large segments in common, their energies are
not
independent. We can approximately take this into account by
assuming a subset of them as independent, leading to  $N\propto
t^{\alpha'}$ for
some $\alpha' < \alpha$. We thus obtain the following estimates for
the mean upper
bound in barrier energy,
\begin{eqnarray}\label{meanE+}
\left\langle E_+(x,t) \right\rangle&=& \left\langle E_c(x,t)
\right\rangle
+\sqrt{2\ln N{\rm var} E_c(x,t)}\nonumber\\
&\simeq& \left( 1 +\beta \sqrt{\ln t} \right)t^{1/3} g_1(x /
t^{2/3}),
\end{eqnarray}
and its variance,
\begin{equation}\label{varE+}
{\rm var}\left( E_+(x,t) \right)={{\rm var}\left( E_c(x,t)
\right)\over 2\ln N}
\simeq {t^{2/3} \over \ln t}g_2(x / t^{2/3}).
\end{equation}
The functions $g_1(y)$ and $g_2(y)$ are proportional to $\sqrt{y}$
and $y$,
respectively, for small $y$,
constant for large $y$, and in general different from that of the
lower bound.

Our numerical simulations indeed confirm the above scaling forms. The
scaling
functions $g_1(y)$ and $g_2(y)$ are plotted in Fig.~\ref{fig1} for
different values
of $t$, after averaging over 2000 realizations of randomness. The
$\sqrt{\ln(t)}$
factors are essential, as a comparable collapse is not obtained
without them.
In fact the best fit to $<E_+(t)>$ is obtained by including the
correction to scaling
factor $4.85 <E_-(t)>$, and with $\beta = 1$. The numerics therefore
support the
neglect of correlations, and the assumption of a gaussian distributed
$E_c(t)$.
As in the lower bound, the initial scaling $\propto\sqrt{ x_f}$ is
not clearly seen
for the sizes studied.  Since the leading power for the scaling of
the lower and upper
bounds is identical, we conclude that the barrier energies also grow
as $t^{1/3}$.
(It remains to be seen if the logarithmic corrections are truly
present, or merely an
artifact of our algorithm.)

We now return to the original question of the response of a flux line
to an external
force,  which in the context of superconductivity is proportional to
the
supercurrent $I$.  The standard
argument\cite{JoffeVinokur,FisherFisherHuse}
assumes an exponential dependence of the net velocity on the {\it
typical barrier
height}.
However, it is quite possible that the overall response of the system
is
determined by the largest, rather than the typical barriers. If so,
knowledge of
the probability distribution of energy barriers is important.
For example, it may be more appropriate to average the {\it waiting
time}
for the activated jumps over  barriers, $\tau\propto\exp(E_B/T)$.
Assuming
a gaussian decay in the tail of the barrier heights, the latter
average,
$[\tau]_{av}\propto\int dE_B e^{E_B/T} P(E_B)$, is dominated by
energies $E_B^* \propto \xi^{\tilde{\psi}}$, with
${\tilde{\psi}}=2\psi=2/3$.
More detailed considerations of such issues
will be taken up in future publications\cite{Mikheev}.

In conclusion, for the simple example of a DPRM, we have
shown that
fluctuations in the minima of the
energy landscape,
and the barriers between them, both scale with the length of the line
as $t^{1/3}$.
It remains to be seen if the upper bound can be further improved
upon, and
placed on a more firm analytical basis. These results provide a
glimpse
into the complexity of the free energy landscape of glassy systems.

This work originated in discussions with Terry Hwa, whose
contribution is
gratefully acknowledged. LVM and MK benefitted from discussions with
other
participants of the ``Vortex Phases" workshop at ITP Santa Barbara
(partly funded
by the NSF grant  number PHY89-04035).  BD is supported  by the
Deutsche
Forschungsgemeinschaft (DFG) under Contract No. Dr 300/1-1. MK
acknowledges
support from NSF grant number DMR-93-03667.

{\bf Note:} After submission of the manuscript we learned of similar
methods
being developed by A. Middleton. One of us (BD) succeeded in
generalizing our algorithm to a FL in 3 dimensions, again confirming
$\psi = \theta$ (see \cite{BD}).

\narrowtext

\end{multicols}


\begin{references}{}
\bibitem{glass1}
K. H. Fischer and J. A. Hertz, {\it Spin glasses}, Cambridge
University Press, Cambridge (1991).

\bibitem{glass2}
J. A. Mydosh, {\it Spin glasses: an experimental introduction},
Taylor \& Francis, London (1993).

\bibitem{expreview}
G. Blatter, M. V. Feigel'man, V. B. Geshkenbein, A. I. Larkin, and V.
M. Vinokur,
Rev.\ Mod.\ Phys.\ {\bf 66}, 1125 (1994).

\bibitem{FisherFisherHuse}
D. S. Fisher, M. P. A. Fisher, and D. A. Huse, Phys.\ Rev.\ B\ {\bf
43}, 130 (1991).

\bibitem{BoseGlass}
D. R. Nelson and V. M. Vinokur, Phys.\ Rev.\ B\ {\bf 48}, 13060
(1993).

\bibitem{JoffeVinokur} L. Ioffe and V. M. Vinokur, J.\ Phys.\ C\ {\bf
20}, 6149 (1987).

\bibitem{KimAnderson}
P. W. Anderson and Y. B. Kim, Rev.\ Mod.\ Phys.\ {\bf 36}, 39 (1964).

\bibitem{HuseHenley}
D. A. Huse and C. L. Henley, Phys.\ Rev.\ Lett.\ {\bf 54}, 2708
(1985).

\bibitem{KardarRev}
M. Kardar, {\it Lectures on Directed Paths in Random Media},  Les
Houches Summer School on Fluctuating Geometries in Statistical
Mechanics
and Field Theory, August 1994 (to be published; see
cond-mat/9411022).

\bibitem{HwaFisher}
T. Hwa and D. S. Fisher, Phys.\ Rev.\ B\ {\bf 49}, 3136 (1994).

\bibitem{footnote1}
The actual choice depends on the specific orientations of the optimal
paths at
the two endpoints. If it is possible to move $(t, x_1(t))$ in the
first elementary move,
the connection from the midpoints is made to $(t, x_2(t))$. If the
move to $(t, x_2(t))$
can be made in the last elementary move, the midway points are
connected to
$(t, x_1(t))$. If neither is possible, subsequent moves determine
this choice.

\bibitem{footnote2}
The astute reader may be worried about the feasibility of moving the
upper and lower
segments independently, given that the midpoint ${(t / 2, x)}$ is
common to both. Depending
on the orientations of the minimal paths arriving at this point, we
found a set of rules on
our discrete lattice by which the midpoint could always be shifted
when switching from
moving one segment to the other. As describing the precise rules is
somewhat lengthy,
we leave the details to future publications.

\bibitem{Mikheev}
L. V. Mikheev, unpublished.

\bibitem{BD} B. Drossel, to be published in J. Stat. Phys. (1995).

\end{references}
\end{document}